\documentclass[letterpaper,twocolumn,10pt]{article}
\usepackage{usenix2019_v3}

\usepackage{amssymb}
\usepackage{tikz}
\usepackage{amsmath}
\usepackage{multirow}
\usepackage{enumitem}
\usepackage{pgfplots}
\pgfplotsset{compat=1.18}

\setlist{nosep, leftmargin=*, topsep=2pt, partopsep=0pt}
\setlist[itemize]{itemsep=1pt}
\setlist[enumerate]{itemsep=1pt}

\newlength{\subsecvspace}
\makeatletter
\renewcommand{\paragraph}{\@startsection{paragraph}{4}{\z@}%
  {1.5ex plus 0.2ex minus 0.1ex}%
  {-1em}%
  {\normalfont\normalsize\bfseries}}
\makeatother

\usetikzlibrary{positioning, backgrounds, fit}

\begin{document}

\date{}

\title{\Large \bf From ASIC to Fleet: \\
  Lessons from Building and Operating a Hyperscaler NIC
  }



\author{
{\rm Prankur Gupta, Alexander Duyck, Jakub Kicinski, Joseph Provine, Neal Peacock, }\\
{\rm Prabhakaran Ganesan, Rajiv Krishnamurthy, Chen Liu, Akshay Viswakumar, Timothy Vitkin,}\\
{\rm Jie Meng, Beatriz Padilla Hernandez, Michael Edwards, Andrei Kozlov, Viren Nathan,}\\
{\rm Tianyi Cui, Joy Chaoyue Xiong, Raul Hormazabal, Mohsin Bashir, Fred Feng, Nathan Walker, }\\
{\rm Lavin Khandelwal, Matt Maia, Milo Piazza, Mohanraj Thillainayagam, Mahmoud Mehr,}\\[3pt]
{\rm Amithash Prasad, and Lee Trager}\\[3pt]
Meta Platforms, Inc.
}

\maketitle
\pagestyle{empty}
\thispagestyle{empty}

\begin{abstract}
We describe the operational infrastructure built to deploy and
operate fbnic, a custom multi-host NIC, across hundreds of
thousands of production hosts at Meta. Vendor multi-host NICs, designed by retrofitting single-host
architectures, suffered from shared firmware and buffers that
created cascading isolation failures over seven years. fbnic
eliminates these through physical isolation, but shifting to
in-house hardware shifts the entire operational burden to the
hyperscaler.

We present a hardware-in-the-loop CI pipeline testing firmware,
driver, and kernel cross-products; a unified observability pipeline
co-locating NIC and switch counters for cross-layer fault
attribution; a driver-first architecture with fewer than ten
firmware message types; a targeted firmware upgrade orchestrator at sub-sled
granularity; and scoped repair automation confining blast radius to
individual host slices. Over ten months, fbnic achieved a
12$\times$ reduction in unplanned unavailability, 37\% lower mean
time to repair, and 2.3$\times$ fewer hardware swaps compared to
vendor NICs on the same platform.
\end{abstract}

\vspace{-10pt}
\section{Introduction}
\vspace{\subsecvspace}
The network interface controller is among the more
operationally consequential components in a hyperscale data center.
Every byte of traffic between hosts traverses a NIC, and every
management operation, including provisioning, firmware updates, health
monitoring, and repair, touches NIC infrastructure. When a NIC fails or
misbehaves, the blast radius extends beyond the individual host: it
can disrupt network-dependent services, stall provisioning workflows,
and, in multi-host architectures where a single NIC is shared across
multiple server blades, take down several hosts simultaneously.

Meta has deployed multi-host NICs (MH-NICs) in every generation of
its Yosemite server platform since 2015~\cite{meta-server-refresh-2017}, motivated initially by
switch port density savings and later by power and cost efficiency.
However, vendor MH-NICs were designed by retrofitting conventional
single-host NIC architectures to support multiple hosts, inheriting
shared firmware, shared packet buffers, and shared management
connectivity. This architectural shortcoming led to a series of
production incidents in which a fault on one host propagated to its
neighbors: a firmware hang could render all four hosts on a sled
unreachable, a congested host's backpressure could degrade
throughput for co-located hosts, and in one case, a NIC failure
made the baseboard management controller (BMC) completely
inaccessible, requiring physical sled removal to recover. These
isolation failures, combined with vendor firmware release cycles
that stretched fixes across six to twelve months, motivated Meta
to develop fbnic, a custom NIC ASIC designed from the ground up
for multi-host operation with complete physical isolation between
hosts.

Building the ASIC, however, was only half the challenge. A custom
NIC shifts the entire operational burden, including firmware development,
testing, qualification, fleet deployment, monitoring, diagnosis, and
repair, from the vendor to the hyperscaler. Vendor NICs arrive with
established firmware release processes, pre-built diagnostic
tooling, and years of fleet-hardened operational playbooks. A
first-generation custom NIC has none of these. It must intercept new
server platforms that are themselves under development, qualify
against a combinatorial matrix of platform component firmware
versions (BMC, BIOS, retimer) and rack switch vendors, and then scale from lab to fleet while
maintaining the reliability standards that production services
demand. At the time of writing, fbnic has scaled from an initial
deployment of a few thousand hosts to hundreds of thousands of
production hosts in under a year, with plans to scale to millions.

This paper describes the operational infrastructure we built to
deploy and operate fbnic at scale across Meta's fleet. We focus
on the systems and processes that sit between the ASIC and production
traffic: the testing pipeline that validates every firmware release,
the observability architecture that enables cross-layer fault
attribution, the repair automation that scopes blast radius to
individual host slices, and the firmware rollout orchestrator that
can upgrade NIC firmware at the per-NIC level, draining only the
hosts sharing the target NIC rather than the entire sled. These systems are not specific to fbnic's ASIC
design; they address challenges that any organization operating
custom network hardware at hyperscale will face.

Meta has operated multi-host NICs at scale since
2015~\cite{meta-server-refresh-2017, meta-accel-infra-2019}, and the
architecture is increasingly mainstream. The Open
Compute Project's DC-MHS (Data Center Modular Hardware System)
specification, co-developed by Intel, Meta, Microsoft, HPE, and
Dell, standardizes multi-host server chassis with shared NIC and
I/O modules, and multiple OEMs now ship DC-MHS
platforms~\cite{ocp-dcmhs}. More broadly, AI workloads are driving
rack density beyond 100\,kW~\cite{afcom-density}, making shared
infrastructure across multiple host modules a design goal. Although AI back-end fabrics demand ever-higher network-to-compute ratios
served by dedicated per-accelerator NICs, this reinforces rather than
undermines the case for a shared front-end NIC: densifying the
general-purpose fleet that AI infrastructure relies on frees rack space
and power for the back-end. Cloud providers face analogous isolation challenges in their multi-tenant SmartNIC and DPU
architectures~\cite{accelnet}. The operational lessons in this
paper generalize to any custom hardware component entering a
hyperscaler fleet.

We make the following contributions:

\begin{itemize}
\item A \textbf{hardware-in-the-loop CI pipeline} that tests the
  cross-product of firmware, driver, and kernel versions daily,
  including seven multi-host isolation tests derived from production
  incidents and multi-day runs under production-derived
  traffic profiles (\S\ref{sec:testing}).

\item A \textbf{unified NIC observability pipeline}, built on
  existing host telemetry infrastructure~\cite{evolvable-telemetry},
  that extends it with NIC hardware counters co-located with matched
  rack switch port counters in a single per-host sample, enabling
  cross-layer fault attribution at fleet scale
  (\S\ref{subsec:observability}).

\item A \textbf{targeted firmware upgrade orchestrator} that
  enables firmware upgrades scoped to the hosts sharing a single
  NIC rather than the entire sled, reducing fleet-wide rollout time
  from approximately 60~days to under four weeks
  (\S\ref{subsec:fw-rollout}).

\item \textbf{Scoped repair automation} that leverages fbnic's
  physical isolation architecture to confine repair blast radius to
  individual host slices, with escalation to broader scope only
  when shared infrastructure is implicated
  (\S\ref{subsec:serviceability}).

\item \textbf{Production incident analyses}, including a
  firmware DMA race condition that caused silent memory
  corruption, with quantitative data covering
  reliability, firmware rollout velocity, and fleet scaling
  (\S\ref{sec:deployment}).
\end{itemize}

\vspace{-6pt}
\section{Background and Motivation}
\label{sec:architecture}
\vspace{\subsecvspace}

\subsection{Multi-Host NICs at Meta}
\label{subsec:mh-nic-history}
\vspace{\subsecvspace}

Meta has deployed multi-host NICs, where a single physical NIC is
shared across multiple server blades in a sled, in every generation
of its Yosemite server platform since
2015~\cite{meta-server-refresh-2017, meta-accel-infra-2019}
(Table~\ref{tab:mh-nic-generations}). The original motivation was
switch port density; subsequent generations added power savings and
future-proofed per-host bandwidth. The latest generation supports
two server tiers with different compute and memory configurations
(Tier~1 and Tier~2, abbreviated T1 and T2).

\begin{table}[t]
\centering
\small
\begin{tabular}{lrrr}
\hline
\textbf{Generation} & \textbf{Hosts/NIC} & \textbf{Link} & \textbf{Per-Host BW} \\
\hline
YV1 (2015) & 4 & 50G  & 12.5G \\
YV2 (2017) & 4 & 50G  & 12.5G \\
YV3 (2019) & 4 & 100G & 25G \\
YV4 (2024) & 2 & 200G & 100G \\
\hline
\end{tabular}
\caption{Multi-host NIC evolution across Yosemite platform
generations. Host count per NIC decreased while per-host bandwidth
increased 8$\times$. Meta's use of multi-host NICs across these
generations has been described in prior public
disclosures~\cite{meta-server-refresh-2017, meta-accel-infra-2019,
ocpus18-mhnic, ocp2020-mhnic, fbnic-ocp}.}
\label{tab:mh-nic-generations}
\end{table}

However, the vendor NICs used in these platforms were not designed
for multi-host operation. They were conventional single-host NICs
modified to support multiple PCIe interfaces, inheriting shared
firmware, shared packet buffers, and shared management
connectivity~\cite{Stephens2019LoomFA}. This architectural
shortcoming created a cascade of isolation failures and
increasingly complex workarounds that ultimately led Meta to
build fbnic~\cite{fbnic-ocp}.

\vspace{-8pt}
\subsection{The Isolation Problem}
\label{subsec:isolation-problem}
\vspace{\subsecvspace}

The difficulty in achieving true isolation in vendor multi-host NICs
stems from three inherent tradeoffs in the shared
hardware~\cite{kong-nsdi23, GrantFairNic, Stephens2019LoomFA}:

\begin{itemize}
    \item \textbf{Shared buffers and flow control.} Vendor MH-NICs
    used a single shared packet buffer (as small as 512\,KB) across
    all hosts. A slow receiver or traffic microburst on one host
    could saturate the buffer, causing packet loss for all hosts on
    the sled. The standard mitigation, enabling IEEE~802.3x pause
    frames to push backpressure to the rack switch's larger
    buffer, traded one problem for several others, as we describe
    below.

    \item \textbf{Microarchitecture resource contention.} Even with
    logically separate queues, hosts compete for DMA engines, PCIe
    bandwidth, and the packet processing pipeline. Hardware arbiters
    (round-robin, weighted-fair-queueing) are too coarse to prevent
    head-of-line blocking under realistic workload
    mixes~\cite{kong-nsdi23, Stephens2019LoomFA}.

    \item \textbf{Per-host resource isolation on the NIC.} Pinning
    dedicated hardware descriptor rings to each host prevents one
    host from monopolizing NIC resources, but faces a scalability
    wall: on-chip descriptor storage is limited, and as host count
    increases the NIC must swap ring state to host memory, causing
    cache thrashing that spikes latency~\cite{kong-nsdi23}.
\end{itemize}

\vspace{-6pt}
\subsection{The Pause Frame Cascade}
\label{subsec:pause-cascade}
\vspace{\subsecvspace}

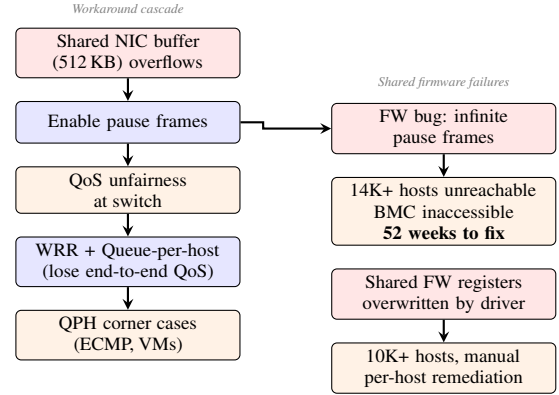
\begin{figure}[t]
\centering
\scriptsize
\begin{tikzpicture}[
    node distance=0.3cm and 0.1cm,
    box/.style={draw, rounded corners=2pt, minimum height=0.55cm,
                text width=2.8cm, align=center, font=\scriptsize},
    problem/.style={box, fill=red!10},
    fix/.style={box, fill=blue!10},
    consequence/.style={box, fill=orange!10},
    arrow/.style={->, >=stealth, thick},
]
\node[problem] (p1) {Shared NIC buffer\\(512\,KB) overflows};
\node[fix, below=of p1] (f1) {Enable pause frames};
\node[consequence, below=of f1] (c1) {QoS unfairness\\at switch};
\node[fix, below=of c1] (f2) {WRR + Queue-per-host\\(lose end-to-end QoS)};
\node[consequence, below=of f2] (c2) {QPH corner cases\\(ECMP, VMs)};

\node[problem, right=1.2cm of p1, yshift=-1.0cm] (fw1)
  {FW bug: infinite\\pause frames};
\node[consequence, below=of fw1] (fw1c)
  {14K+ hosts unreachable\\BMC inaccessible\\\textbf{52 weeks to fix}};
\node[problem, below=of fw1c] (fw2)
  {Shared FW registers\\overwritten by driver};
\node[consequence, below=of fw2] (fw2c)
  {10K+ hosts, manual\\per-host remediation};

\draw[arrow] (p1) -- (f1);
\draw[arrow] (f1) -- (c1);
\draw[arrow] (c1) -- (f2);
\draw[arrow] (f2) -- (c2);
\draw[arrow] (f1.east) -- ++(0.3,0) |- (fw1.west);
\draw[arrow] (fw1) -- (fw1c);
\draw[arrow] (fw2) -- (fw2c);

\node[font=\tiny\itshape, gray, above=0.1cm of p1] {Workaround cascade};
\node[font=\tiny\itshape, gray, above=0.1cm of fw1] {Shared firmware failures};
\end{tikzpicture}
\caption{The pause frame cascade on vendor multi-host NICs. Each
workaround (blue) introduced new failure modes (orange). Shared
firmware created the most severe outages (red), requiring weeks to
months for vendor fixes.}
\label{fig:pause-cascade}
\end{figure}

Meta's operational experience with vendor MH-NICs illustrates how
software workarounds for hardware isolation deficiencies create
compounding operational complexity~\cite{ghabashneh-imc22}.
Figure~\ref{fig:pause-cascade} summarizes the cascade: each layer
of mitigation introduced new failure modes that required further
workarounds.

The cascade began when Meta enabled IEEE~802.3x pause frames
fleet-wide to reduce packet loss from the small shared NIC buffer.
This transferred the isolation problem to the rack switch (RSW),
where pause frames caused cross-host QoS unfairness: the switch's
QoS classes operated per-port, not per-host, so bursty
high-priority traffic to one host could starve lower-priority
traffic to all other hosts on the same port. Switching to weighted
round-robin (WRR) sacrificed strict priority guarantees.
Head-of-line blocking within the same priority class led to
\emph{queue-per-host} (QPH) in the switch software: the RSW
allocated a separate hardware queue for each host behind the
shared NIC port, bandwidth-limiting each queue to the host's
maximum achievable rate. QPH consumed all available QoS classes
on the RSW-to-NIC link for per-host isolation, eliminating
end-to-end traffic differentiation.

QPH itself proved fragile: ECMP routes broke per-host queue
assignment for 552~racks, and virtual machine MAC addresses
confused the algorithm, re-creating the noisy-neighbor problem.
Each corner case required months of engineering to resolve.

The most severe consequences involved shared firmware. A vendor
firmware bug caused NICs to issue pause frames indefinitely in
response to PCIe errors on a single host, making all hosts on the
sled and the BMC unreachable. The only recovery was to manually
disable pause frame reception on the affected switch port and
power-cycle the sled. Over 1,500~racks exhibited this failure,
potentially affecting 14,000+~hosts. The vendor required
\textbf{52~weeks} to deliver a fix. A separate firmware bug caused
traffic stalls during neighbor reboots; the vendor stated that the
correct fix was ``not possible'' in their architecture, and the
eventual workaround took 22~months to deploy fleet-wide. In another scenario, a kernel driver update overwrote shared NIC firmware
registers, breaking isolation for over tens of thousands of machines. The
corruption persisted across reboots and kernel rollbacks,
requiring weeks of manual per-host remediation. The fundamental
problem was architectural: every software workaround introduced to
compensate for the lack of hardware isolation created new edge
cases and operational complexity.

\vspace{-6pt}
\subsection{fbnic Architecture}
\label{subsec:fbnic-arch}
\begin{figure}[t]
    \centering
    \includegraphics[width=0.75\linewidth]{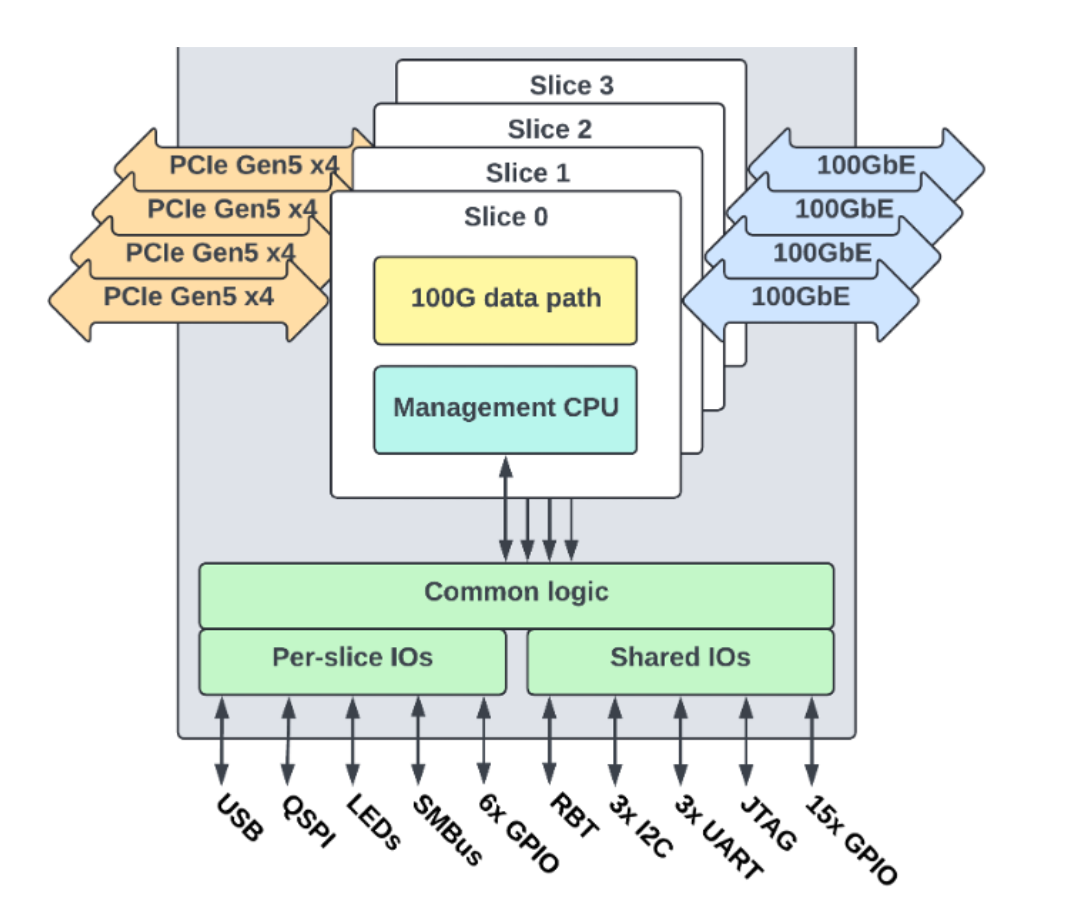}
    \caption{fbnic hardware isolation architecture. Each host
    connects to an independent NIC slice with its own MAC, firmware
    instance, PCIe interface, and packet buffers. Only board power
    and the network-facing transceiver are shared. Architecture
    previously described in~\cite{fbnic-ocp}.}
    \label{fig:fbnic-isolation}
\end{figure}

\vspace{\subsecvspace}

fbnic was designed to eliminate this cascade of workarounds
through \emph{complete physical isolation} at the hardware level
(Figure~\ref{fig:fbnic-isolation}). Each host connects to an
independent slice of the NIC with its own MAC, its own firmware
instance, its own PCIe interface, and its own packet buffers. The
only shared resources are board-level power and the network-facing
transceiver.

This architecture provides three properties that vendor MH-NICs
could not:

\begin{enumerate}
\item \textbf{No pause frame propagation.} Because each host has
  an independent MAC, backpressure from a congested host cannot
  generate pause frames that affect its neighbors' traffic. The
  QPH workaround---and the QoS sacrifice it required---becomes
  unnecessary, re-enabling end-to-end traffic differentiation on
  the RSW-to-NIC link.

\item \textbf{Independent failure domains.} Vendor MH-NICs run a
  single firmware instance shared across all hosts; a firmware bug,
  hang, or hardware fault on one host can corrupt state or halt
  traffic for all hosts on the sled
  (\S\ref{subsec:pause-cascade}). fbnic runs an independent
  firmware instance per host slice with no shared state, so a fault
  on one slice, whether firmware, PCIe, or link-level, cannot
  affect the others. This property directly enables scoped repair
  automation (\S\ref{subsec:serviceability}) and sub-sled firmware
  upgrades (\S\ref{subsec:fw-rollout}).

\item \textbf{Independent observability.} Per-host MAC counters
  allow the telemetry pipeline (\S\ref{subsec:observability}) to
  attribute errors to individual host ports, rather than reporting
  aggregate NIC-level statistics that obscure which host is
  affected.
\end{enumerate}

\noindent

\paragraph{Cost and power savings.} Per-host isolation retains the
efficiencies that motivate multi-host NICs: hosts share one
network-facing transceiver, link, and board. Because fbnic's silicon cost is fixed rather than scaled by bandwidth (vendors price by bandwidth), the design saves roughly 28\% per host at 100\,Gbps, 40\% at 200\,Gbps, and 62\% at 400\,Gbps versus an equivalent vendor NIC, widening with bandwidth. It also draws less NIC power than a discrete NIC per host, though at modern CPU TDPs (200--400\,W) this per-NIC delta has become a small fraction of total system power.

\paragraph{The cost of isolation.} Physical isolation is not free.
Dedicating resources per host forgoes the pooling a shared design permits. The per-host datapath (MAC, DMA engines, queue management, and buffers) is replicated across the NIC's four slices on a single 38\,mm$^2$ die (5\,nm)~\cite{fbnic-ocp}, adding die area relative to a shared pipeline. More consequential in production, each host's buffer cannot borrow from an idle neighbor: fbnic's smaller per-host receive buffer (384\,KB, versus a comparable vendor NIC's shared ${\sim}$700\,KB pool) leaves it exposed to a correlated 50--70\,$\mu$s CPU PCIe stall that a shared buffer could absorb (\S\ref{subsec:pcie-latency}). The trade is deliberate: for less statistical-multiplexing headroom under rare stalls, isolation removes cross-host buffer contention in steady state, where production receive-buffer exhaustion is near zero on fbnic versus thousands on the vendor NIC (\S\ref{subsec:eval-reliability}).

\section{Design for Operations}
\label{sec:design-for-ops}
\vspace{\subsecvspace}

The physical isolation architecture described in
\S\ref{subsec:fbnic-arch} directly informed three operational
design decisions. First, a \emph{driver-first} configuration
minimizes firmware complexity by placing data-path logic in the
kernel driver rather than in opaque firmware
(\S\ref{subsec:driver-first}). Second, serviceability
infrastructure leverages per-host failure domains for scoped
repair and automated two-channel diagnostics
(\S\ref{subsec:serviceability}). Third, an observability pipeline
exploits per-host MAC counters to co-locate NIC and switch
telemetry for cross-layer fault attribution
(\S\ref{subsec:observability}).

\subsection{Driver-First Configuration}
\label{subsec:driver-first}
\vspace{\subsecvspace}

A key architectural decision in fbnic was to place nearly all
data-path configuration logic in the Linux kernel driver rather
than in NIC firmware~\cite{fbnic-netdev}. This \emph{driver-first} model contrasts
sharply with vendor NICs, where firmware acts as an opaque
intermediary that owns most hardware configuration: the driver
sends commands to firmware, and firmware programs the hardware on
the driver's behalf. In a driver-first model, the driver programs
hardware registers directly, and firmware is restricted to a
narrow management-plane role.

Table~\ref{tab:driver-fw} summarizes the division for runtime
operation. The fbnic firmware initially configures the NIC for
BMC communication before the driver loads; once the driver takes
ownership, it reprograms all data-path state directly. The
runtime firmware interface consists of three request/response
pairs: capability advertisement, ownership handshake, and periodic
heartbeat, with a small number of additional commands for PHY
configuration. By contrast, a representative vendor NIC defines
over 170~Ethernet firmware command opcodes in its upstream driver
interface header (excluding RDMA-specific and debug commands).

\begin{table}[t]
\centering
\small
\begin{tabular}{lcc}
\hline
\textbf{Function} & \textbf{Vendor} & \textbf{fbnic} \\
 & \textbf{(FW-first)} & \textbf{(Driver-first)} \\
\hline
Queue setup / teardown  & FW cmd  & Driver $\rightarrow$ HW \\
MAC address filters     & FW cmd  & Driver $\rightarrow$ TCAM \\
RSS hash \& indirection & FW cmd  & Driver $\rightarrow$ HW \\
Link speed \& FEC       & FW owns & Driver $\rightarrow$ PCS \\
Statistics collection   & FW      & Driver $\rightarrow$ HW \\
Runtime FW commands     & 170+    & \textbf{$<$10} \\
\hline
\end{tabular}
\caption{Runtime configuration responsibility: vendor firmware-first
vs.\ fbnic driver-first. ``Driver $\rightarrow$ HW'' indicates the
driver programs hardware registers directly without firmware
involvement.}
\label{tab:driver-fw}
\end{table}

This design has four operational consequences. First, it
\textbf{shrinks the firmware attack surface}. The fewer functions
firmware performs, the fewer firmware bugs can occur. The memory
corruption incident described in \S\ref{sec:incidents}, a race
condition in the firmware's DMA address translation logic, occurred
in one of the few remaining firmware-owned code paths. With a
firmware-first architecture, the surface area for such bugs would
be larger.

Second, it \textbf{accelerates fix velocity for data-path issues}.
Because the driver owns queue management, packet classification,
and link configuration, bugs in these functions are fixed through
kernel updates, which can be deployed via live patches within
days for critical issues, rather than through the firmware
qualification pipeline described in \S\ref{sec:testing}, which
requires hardware-in-the-loop testing and takes approximately two
weeks. Vendor NIC data-path bugs, by contrast, require firmware
fixes with qualification cycles measured in months
(\S\ref{subsec:pause-cascade}).

Third, it provides a \textbf{transparent and auditable control
plane}. The entire data-path configuration is implemented in
open-source kernel code (the fbnic driver was merged into
upstream Linux v6.11+, with the full source available for
community review and independent validation), visible to standard
kernel tracing tools.
When a NIC misbehaves, engineers can trace the exact register
writes that configured it, rather than filing a support request
with a vendor to inspect opaque firmware state.

Fourth, it \textbf{avoids capability ossification}. A firmware-first NIC
exposes only the features its firmware implements, limiting the driver to
what the vendor chose to surface; programming the hardware directly lets us
use the silicon's full capability and enable new features without waiting
on a firmware release.

\paragraph{Performance considerations.} Driver-first is a decision about
\emph{where configuration logic runs}, not about the data path. The driver
programs the same hardware offloads directly, including checksum, TCP
segmentation, generic segmentation and receive offload, hardware
traffic-control (qdisc) offload, and earliest-departure-time rate limiting,
so bulk throughput and per-packet processing remain fully
hardware-accelerated. The performance-relevant difference is confined to
the control plane: configuration operations such as queue setup, filter
programming, and link training are direct register writes rather than
driver-to-firmware round-trips, reducing configuration latency at the cost
of additional driver complexity. This yields no data-path penalty in
production: on identical services and platforms, fbnic matches or beats
vendor NIC TCP retransmit rates with near-zero receive-buffer exhaustion
(Table~\ref{tab:perf-parity}).

\subsection{Serviceability from Day One}
\label{subsec:serviceability}
\vspace{\subsecvspace}

Operating NICs at fleet scale means that failures are not exceptions
but statistical certainties. One design goal for fbnic was to
minimize the \emph{mean time to resolution}, which we define as
the elapsed time from identifying a bug to rolling out its fix across
the fleet. With vendor NICs, this cycle was dominated by external
dependencies: diagnosing an issue often required vendor support
interactions with limited visibility into firmware internals, and
once a fix was produced, vendor firmware release cycles and subsequent
internal validation extended the timeline to several months before a
patch reached even a fraction of the fleet. We designed fbnic's
serviceability around three principles: own the fix cycle, automate
evidence collection, and scope repair blast radius to the minimum
necessary.

\paragraph{Owning the Fix Cycle.}
Because fbnic firmware and drivers are developed in-house, bug
fixes follow the same continuous integration and deployment pipeline
used for other infrastructure software. A firmware change is
validated through the two-stage CI pipeline
(\S\ref{subsec:lab-ci}) and deployed fleet-wide in approximately
five weeks. Vendor NIC fixes historically required months, as
quantified in \S\ref{subsec:eval-velocity}.

\paragraph{Automated Diagnostic Log Collection.}
Effective root-cause analysis depends on capturing diagnostic data
at the time of failure, before a host is reimaged or a NIC is
replaced. For fbnic, we built automated log collection that is
triggered whenever a NIC-related repair event is initiated,
collecting diagnostics through two independent channels and
attaching them directly to the repair ticket. \emph{In-band
collection} runs on the host and gathers NIC register dumps,
driver metadata, interface statistics, firmware logs, and
host-level context (kernel logs, PCI topology, interrupt state).
\emph{Out-of-band collection} operates via the BMC, independent of
host availability, capturing firmware console logs, PCIe link
state, and PHY status. Out-of-band collection is meaningful for multi-host NICs: when one host is inaccessible, the
BMC path remains available because it is shared across the NIC
rather than tied to any individual host.

\noindent
The two-channel approach ensures that diagnostic evidence is
preserved even in scenarios where the in-band path is unavailable
(e.g., a host that fails to boot or a crashed firmware slice),
significantly improving the quality of post-mortem analysis.

\paragraph{Scoped Repair Automation.}
fbnic's physical isolation (\S\ref{sec:architecture}) enables
repair automation that targets only the affected host's NIC slice,
leaving other hosts on the same physical NIC undisturbed. The
system escalates to a full NIC-level repair only when diagnostic
evidence indicates a problem with shared physical infrastructure
(board-level power or the network-facing transceiver).

\subsection{Observability Architecture}
\label{subsec:observability}
\vspace{\subsecvspace}

A recurring lesson from operating vendor NICs was that NIC-level
observability was inadequate for production triage. NIC health
signals, including FEC error rates, buffer exhaustion events, link flap
counts, were collected as individual per-host time series, siloed
from one another and from switch-side metrics. There was no unified
dataset correlating NIC counters with the corresponding rack switch
port, and no way to slice NIC health by service, workload, firmware
version, or server type. Diagnosing whether the root cause of
elevated latency was the NIC, the network, or the host required
manually SSH-ing into individual hosts, inspecting \texttt{ethtool}
counters, and cross-referencing switch-side data from a separate
tool, an approach that did not scale. For fbnic, we designed
observability as a requirement from the outset.

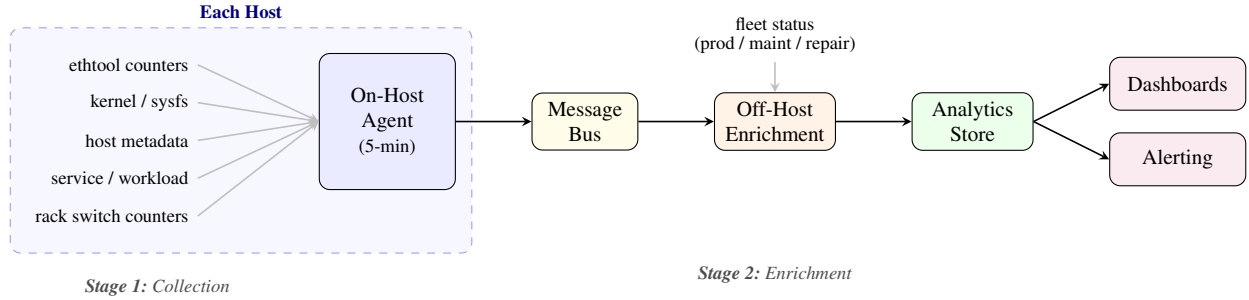
\begin{figure*}[t]
\centering
\begin{tikzpicture}[
    node distance=0.4cm,
    box/.style={draw, rounded corners, minimum height=0.7cm,
                minimum width=1.6cm, align=center, font=\footnotesize},
    agentbox/.style={box, fill=blue!8, minimum height=1.8cm,
                     minimum width=1.8cm},
    pipebox/.style={box, fill=yellow!10, minimum width=1.4cm},
    taggerbox/.style={box, fill=orange!10, minimum width=1.6cm},
    storebox/.style={box, fill=green!8, minimum width=1.6cm},
    dashbox/.style={box, fill=purple!8, minimum width=1.8cm},
    src/.style={font=\scriptsize, anchor=east},
    arrow/.style={->, >=stealth, semithick},
    stagelabel/.style={font=\scriptsize\itshape, text=gray!60!black},
]

\node[agentbox] (agent) {On-Host\\Agent\\{\scriptsize (5-min)}};

\node[src, left=1.6cm of agent, yshift=0.75cm]  (s1) {ethtool counters};
\node[src, left=1.6cm of agent, yshift=0.25cm]  (s2) {kernel / sysfs};
\node[src, left=1.6cm of agent, yshift=-0.25cm] (s3) {host metadata};
\node[src, left=1.6cm of agent, yshift=-0.75cm] (s4) {service / workload};
\node[src, left=1.6cm of agent, yshift=-1.25cm] (s5) {rack switch counters};

\foreach \src in {s1, s2, s3, s4, s5}
    \draw[arrow, gray!50] (\src.east) -- (agent.west);

\begin{scope}[on background layer]
\node[draw=blue!40, dashed, rounded corners=4pt, fill=blue!3,
      fit=(s1)(s5)(agent), inner xsep=6pt, inner ysep=8pt,
      label={[font=\scriptsize\bfseries, blue!50!black]above:Each Host}] {};
\end{scope}

\node[pipebox, right=1.0cm of agent] (bus) {Message\\Bus};
\draw[arrow] (agent) -- (bus);

\node[taggerbox, right=1.0cm of bus] (tagger) {Off-Host\\Enrichment};
\draw[arrow] (bus) -- (tagger);

\node[font=\scriptsize, above=0.4cm of tagger, align=center] (fs)
    {fleet status\\[-1pt]{\scriptsize (prod / maint / repair)}};
\draw[arrow, gray!50] (fs) -- (tagger);

\node[storebox, right=1.0cm of tagger] (store) {Analytics\\Store};
\draw[arrow] (tagger) -- (store);

\node[dashbox, right=1.0cm of store, yshift=0.5cm] (dash)
    {Dashboards};
\node[dashbox, right=1.0cm of store, yshift=-0.5cm] (alert)
    {Alerting};
\draw[arrow] (store.east) -- (dash.west);
\draw[arrow] (store.east) -- (alert.west);

\node[stagelabel, below=0.5cm of s5, xshift=0.6cm]
    {\textbf{Stage 1:} Collection};
\node[stagelabel, below=1.4cm of tagger]
    {\textbf{Stage 2:} Enrichment};

\end{tikzpicture}
\caption{Overview of the NIC observability pipeline, from on-host
collection through off-host enrichment to the analytics store.}
\label{fig:observability-pipeline}
\end{figure*}

\paragraph{Pipeline Architecture.}
We built a two-stage telemetry pipeline to collect, enrich, and analyze
NIC health data fleet-wide (Figure~\ref{fig:observability-pipeline}). The first stage runs on every host as a
dedicated thread within an existing host-level daemon. Every five minutes,
it reads NIC counters from the driver via \texttt{ethtool} and kernel
interfaces, along with host metadata (server type, NIC firmware version,
kernel version, BIOS revision). Importantly, the on-host agent also
queries the attached rack switch directly: it discovers the switch
identity and port mapping via a network topology service, then fetches
port-level counters (e.g., interface flaps, input/output discards,
FEC errors, buffer watermarks) from the switch's management interface.
This co-locates NIC-side and switch-side counter collection in the same
sampling interval, ensuring temporal alignment between the two ends of
the link. The collected record is emitted to a distributed message bus.

The second stage is an off-host enrichment service that consumes these
records and augments them with data not efficiently obtainable on-host,
such as the host's current fleet status (e.g., whether it is in
production, undergoing maintenance, or in repair).
The enriched records are written to a time-series analytics store,
enabling near-real-time queries with arbitrary grouping and filtering
across the fleet.

\paragraph{Signal Selection and Operational Use.}
Table~\ref{tab:observability-signals} summarizes the signals
collected per sample and their operational use. By joining NIC-side
and switch-side metrics in the same record, the pipeline enables
\emph{cross-layer fault attribution}: determining whether TCP
retransmits are caused by NIC-level buffer exhaustion or switch-side
congestion discards. This proved important during fbnic
deployment, as the team needed to compare fbnic behavior
side-by-side with vendor NICs running identical workloads to avoid
falsely attributing pre-existing platform issues to the new NIC.
The pipeline also provides per-service dashboards, firmware version
tracking, and rollout regression detection
(\S\ref{sec:deployment}).

\begin{table}[t]
\centering
\small
\begin{tabular}{p{1.1cm}p{1.6cm}p{2.0cm}p{1.8cm}}
\hline
\textbf{Stage} & \textbf{Source} & \textbf{Signals} & \textbf{Use} \\
\hline
On-host & ethtool / sysfs & FEC rates, buffer exhaust., pause frames, link events & NIC health \\
On-host & RSW mgmt & Flaps, discards, watermarks, FEC & Cross-layer correlation \\
On-host & Host meta & FW ver., kernel, service, workload & Per-service analysis \\
Off-host & Fleet sys. & Prod / maint / repair status & Alerting, filtering \\
On-host & Config & Coalescing, ring sizes, qdiscs & Drift detection \\
\hline
\end{tabular}
\caption{Observability pipeline signal categories. NIC and switch
counters are co-located in a single per-host sample every 5~min.}
\label{tab:observability-signals}
\end{table}

For fbnic specifically, we added per-interface counters
(e.g., per-MAC receive bytes, per-MAC errors) that leverage the
physical isolation architecture (\S\ref{subsec:serviceability}) to
attribute issues to individual host ports independently.


\section{Testing Infrastructure}
\label{sec:testing}
\vspace{\subsecvspace}
Validating a custom NIC differs from qualifying a vendor NIC: with no
external vendor validating the silicon for us, that burden shifts
in-house. Meta's internal silicon team, with its hardware partners,
handles silicon-level validation of the ASIC (pre-silicon verification on emulation and post-silicon bring-up). This paper focuses on the layer
above: system- and fleet-level validation, where every firmware release
must pass the full matrix of platform configurations before production.
fbnic compounded this challenge because it intercepted new server
platforms simultaneously, each bringing a new BMC codebase, a PCIe
retimer (added because fbnic targeted next-generation PCIe speeds
while the initial platform supported only the prior generation), and
multiple rack switch vendors. The result was a combinatorial test
matrix that could not be covered by conventional NIC qualification
workflows.

We built a hardware-in-the-loop continuous integration pipeline
with 58~hardware test cases spanning power cycling, link stability,
traffic isolation, production workload profiling, PXE boot, platform
compatibility, and upstream kernel self-tests, designed to catch the
classes of issues that are unique to operating a custom multi-host
NIC at fleet scale.

\subsection{Lab Infrastructure and CI Pipeline}
\label{subsec:lab-ci}
\vspace{\subsecvspace}

We repurposed an existing rack switch test lab
(NetCastle~\cite{netcastle}) and extended it with host-level
testing capabilities. One requirement was topology-aware host
reservation: because multi-host NIC tests need to control which
hosts share a physical NIC, the reservation system had to support
reserving hosts across sleds, within a single sled, and across
racks for traffic-pattern tests. We built orchestration services
on top of the existing lab framework to provide this granularity,
managing 16~host pools spanning CI/CD, stress testing, per-platform
qualification (T1, T2), and sandboxing.

Owning both firmware and the kernel driver means that regressions
can originate in either component or in their interaction with a
specific kernel version, and Meta's fleet runs multiple kernel
versions simultaneously. Firmware validation therefore requires
testing the \emph{cross-product} of firmware, driver, and kernel
versions, not just a single combination.

We address this with a two-stage Conveyor pipeline. Stage~1 runs 
\emph{daily} with shorter test durations, optimized for fast 
feedback to developers. It executes five parallel pipelines, each
pinning one axis of the matrix to isolate regressions:
  
\begin{itemize}
\item \textbf{Stable FW + tip-of-tree driver:} catches driver
  regressions against known-good firmware.
\item \textbf{Stable driver/kernel + tip-of-tree FW:} catches
  firmware regressions against known-good host software.
\item \textbf{Three production kernel versions + tip-of-tree FW:}
  catches kernel-specific interactions, since the fleet runs
  multiple kernel versions concurrently.
\end{itemize}

\noindent
Each Stage~1 pipeline runs the following test sequence: install
verification (PCIe enumeration, driver load, link-up), functional
tests (power cycling, link flap recovery, sensor validation, PXE
boot), the multi-host isolation suite
(\S\ref{subsec:multihost-tests}), production traffic profiles
(\S\ref{subsec:traffic-profiles}), upstream kernel selftests
from the \texttt{drivers/net} collection, and Linux Network
Device Tests (LNDT)~\cite{lndt}, with a
version-aware blocklist for known upstream issues unrelated to the
fbnic driver.

Stage~2 runs \emph{weekly} with multi-day test durations, providing
release confidence for firmware candidates that passed Stage~1. It
executes two pipelines, one per major production kernel
version, against the Stage~1-qualified firmware, running extended
workload profiling and disruption suites on per-tier host pools
(T1~and~T2). The longer durations are critical for catching
time-dependent issues such as memory leaks, counter wraparound bugs,
and intermittent link instability that only manifest under sustained
load. A firmware release requires passing both stages before it is
eligible for fleet deployment (\S\ref{subsec:fw-rollout}).

\subsection{Multi-Host Isolation Tests}
\label{subsec:multihost-tests}
\vspace{\subsecvspace}

Validating fbnic's hardware isolation guarantees was the central goal of our test infrastructure: these tests were designed directly from production incidents observed with vendor multi-host NICs, where activity on one host caused performance
degradation on hosts sharing the same NIC.

Each test follows the same pattern: establish baseline iperf traffic
at line rate between a pair of hosts sharing a NIC, then introduce a
disruptive workload on the neighboring hosts sharing that same NIC,
and verify that the undisturbed hosts maintain throughput above 90\%
of line rate with zero pause frame leakage. Seven disruption
scenarios are tested:

\begin{enumerate}
\item \textbf{AC power cycling} of a neighboring host during active traffic.
\item \textbf{DC power cycling} of a neighboring host during active traffic.
\item \textbf{Reprovisioning:} a full PXE boot and OS reprovision on a
  neighboring host.
\item \textbf{XDP churn:} repeated load and unload of XDP programs on a
  neighboring host during active traffic.
\item \textbf{NIC reconfiguration:} queue-count and receive ring-buffer
  changes via \texttt{ethtool} on a neighboring host during active
  traffic.
\item \textbf{Host-induced packet drops with pause-frame verification:}
  saturate a neighboring host (queue count~1, receive ring 128~entries)
  and verify that (a)~undisturbed hosts hold line-rate throughput and
  (b)~the congested host's pause frames do not propagate to undisturbed
  hosts.
\item \textbf{Incast:} drive incast from three external hosts to a single
  fbnic host while undisturbed hosts on the same NIC maintain independent
  flows.
\end{enumerate}

\noindent
All seven scenarios pass consistently across firmware releases:
undisturbed hosts sustain 92.7\,Gbps on 100\,Gbps links (the
7.3\% gap is TCP/IP protocol overhead, not isolation loss). In the
more demanding test, host-induced packet drops with pause frames
enabled, the congested host's receiver accumulates over a million
pause frames while undisturbed hosts on the same physical NIC
observe fewer than 60, confirming that fbnic's per-host MAC
isolation extends to flow control backpressure.
These results validate that fbnic's physical isolation
architecture (\S\ref{sec:architecture}) delivers the per-host
independence that vendor multi-host NICs could not guarantee.

\subsection{Production Traffic Profiles}
\label{subsec:traffic-profiles}
\vspace{\subsecvspace}

Standard benchmarks such as iperf and netperf stress bulk throughput
but do not exercise the interaction patterns that dominate production
traffic. To close this gap, we built synthetic traffic profiles
using uPerf, parameterized from production service telemetry.

We derived four profiles representing distinct workload archetypes:

\begin{itemize}
\item \textbf{Read-read-write} (caching): two 350\,B reads
  followed by a 1300\,B write.
\item \textbf{Continuous small reads} (metadata): 20\,B
  payloads at maximum rate, stressing packets-per-second and
  interrupt coalescing.
\item \textbf{4\,KB read/write} (storage): medium-payload
  bidirectional transfers exercising DMA engine efficiency.
\item \textbf{Read-read-read-write} (mixed): asymmetric
  read-heavy pattern with larger aggregate payload.
\end{itemize}

\noindent
Each profile runs 2000~concurrent threads against three servers, in
both single-client and dual-client (two hosts sharing a NIC)
configurations, yielding eight test variants per firmware candidate.
Beyond throughput, we collect retransmission rates, FEC error
counters, pause frame activity, ECN counters, rack switch buffer
watermarks, and CPU utilization.

Table~\ref{tab:uperf-results} shows results from a representative
Stage~2 run across both platform tiers. Each profile ran for
24~hours except \emph{4\,KB R/W}, which completes in 1--3~hours
due to larger payloads. Across all profiles and tiers, the NIC
sustained zero FEC uncorrectable errors and zero dropped frames.

\begin{table}[t]
\centering
\small
\begin{tabular}{llrrrrr}
\hline
\textbf{Tier} & \textbf{Profile} & \textbf{Dur.} & \textbf{Tput} & \textbf{Rxmit/s} & \textbf{FEC} & \textbf{Drops} \\
 & & (h) & (Gbps) & (client) & (uncorr.) & \\
\hline
T1 & RRW      & 24   & 125.3 & 3.4   & 0 & 0 \\
T1 & R\_20B   & 24   & 62.8  & 1.0   & 0 & 0 \\
T1 & 4K R/W   & 2.9  & 180.1 & 127.2 & 0 & 0 \\
T1 & RRRW     & 24   & 146.7 & 3.4   & 0 & 0 \\
\hline
T2 & RRW      & 24   & 145.2 & 0.4   & 0 & 0 \\
T2 & R\_20B   & 24   & 80.2  & 4.9   & 0 & 0 \\
T2 & 4K R/W   & 0.6  & 179.1 & 274.3 & 0 & 0 \\
T2 & RRRW     & 24   & 147.2 & 0.7   & 0 & 0 \\
\hline
\end{tabular}
\caption{Production traffic profile results (single-client, Stage~2).
Throughput is aggregate bidirectional. FEC uncorrectable errors and
NIC-level dropped frames are zero across all profiles and tiers.}
\label{tab:uperf-results}
\end{table}

\subsection{Platform Compatibility}
\label{subsec:compat-testing}
\vspace{\subsecvspace}

The more persistent source of schedule risk was platform
integration. fbnic intercepted platforms with changing BMC
codebases, PCIe retimers, and multiple rack switch vendors,
creating a ``guilty until proven innocent'' dynamic where platform
issues were attributed to the newest component. In one case,
elevated packet drop rates turned out to be a CPU power management
feature stalling the PCIe bus for 50--70\,$\mu$s
(\S\ref{subsec:pcie-latency}), diagnosed by correlating NIC
buffer drops with PCIe latency across NIC models via the
observability pipeline (\S\ref{subsec:observability}). We run
compatibility tests against every supported combination of BMC
firmware, rack switch vendor, retimer firmware, and BIOS revision,
validating NC-SI communication, out-of-band management, and PXE
boot across the full configuration matrix.

\section{Deployment Experience}
\label{sec:deployment}
\vspace{\subsecvspace}

This section describes two deployment challenges that arose during
fbnic's production ramp: a platform integration issue requiring
cross-vendor collaboration (\S\ref{subsec:pcie-latency}), and the
firmware rollout infrastructure built to address the multi-host
upgrade scope constraint (\S\ref{subsec:fw-rollout}).

\subsection{Platform Integration: The PCIe Latency Problem}
\label{subsec:pcie-latency}
\vspace{\subsecvspace}

The most technically challenging deployment issue fbnic faced was
not a NIC defect but an interaction with the host CPU's power
management. During early production deployment, fbnic hosts
exhibited packet drops in the NIC's receive buffer (FIFO4) that
vendor NICs on the same platform did not. The NIC was immediately
suspected, as it was the newest component, but the root cause was elsewhere, uncovered only after the PCIe retimer, BIOS settings, and CPU power states had each been ruled out.

The host CPU implements a power-saving feature that dynamically
adjusts an internal bus connecting the I/O die to the compute die
to save power. When I/O traffic is low, the bus enters a
reduced-capacity mode; when traffic increases, it returns to full
capacity. The transition
causes PCIe read completions to stall for 50--70\,$\mu$s, far
exceeding the expected 1--2\,$\mu$s latency. fbnic has a
384\,KB receive host buffer, which fills in approximately
30\,$\mu$s at 100\,Gbps line rate. A 50+\,$\mu$s PCIe stall
therefore overflows the buffer, causing packet drops. Vendor NICs
on the same platform were unaffected because they have larger
buffers (${\sim}$700\,KB) that can absorb the stall.

The investigation required cross-functional collaboration spanning
the NIC team, the CPU vendor, and the platform team. After ruling
out the PCIe retimer and CPU power states, increasing PCIe credits
via BIOS settings reduced drops by nearly 100×, confirming a credit-starvation hypothesis. The breakthrough came when disabling
the CPU's internal bus-width folding feature eliminated the drops
entirely.

\paragraph{Solution: Pause frames as a safety net.}
Rather than permanently disabling the CPU's power-saving feature,
which would increase power consumption across hundreds of
thousands of hosts, we enabled TX pause frames on fbnic. When
the NIC's receive buffer approaches capacity, it sends a pause
frame to the rack switch, which temporarily stops sending traffic
until the buffer drains. This is the same mechanism that caused
cascading failures on vendor multi-host NICs
(\S\ref{subsec:pause-cascade}), but with two differences.

First, fbnic's physical isolation ensures that pause frames
from one host's MAC cannot affect traffic to neighboring hosts.
The pause-frame cascade described in \S\ref{subsec:pause-cascade}
---where one host's congestion could starve all hosts on the
sled---is architecturally impossible on fbnic.

Second, pause frames on fbnic are a \emph{rare} event.
Fleet-wide telemetry shows that the median fbnic host sends
\textbf{zero} pause frames per collection interval; even at the
95th percentile, the rate is just 14~frames per interval
(Table~\ref{tab:pause-stats}). The CPU's bus-width transition is
infrequent and brief, so the pause mechanism activates only
during transient PCIe stalls rather than as a sustained traffic
management tool.

\begin{table}[t]
\centering
\small
\begin{tabular}{lr}
\hline
\textbf{Percentile} & \textbf{TX Pause Frames / interval} \\
\hline
P50 (median) & \textbf{0} \\
P95           & 14 \\
P99           & 256 \\
P99.9         & 2,469 \\
\hline
\end{tabular}
\caption{Distribution of TX pause frame rates across the fbnic
fleet. The median host sends zero pause
frames. Contrast with vendor multi-host NICs where pause frames
were the primary traffic management mechanism
(\S\ref{subsec:pause-cascade}).}
\label{tab:pause-stats}
\end{table}

\paragraph{Pause storm protection.}
Enabling pause frames introduced a residual risk: if a host hangs
entirely (CPU lockup, kernel panic), the NIC cannot drain its
receive buffer and will send pause frames indefinitely, potentially
exhausting the rack switch's buffer and affecting other hosts under
the same switch. This is the same failure mode as the vendor NIC
incident in \S\ref{subsec:pause-cascade}, albeit scoped to the
switch rather than the sled.

To bound this risk, we implemented hardware-backed pause storm
protection. If pause frames are asserted continuously for
500\,ms, the NIC hardware automatically disables pause
transmission, allowing packets to drop at the NIC rather than
flooding the switch. This bounds the impact of a host hang to a
single 500\,ms burst of pause frames.

\subsection{Firmware Rollout Infrastructure}
\label{subsec:fw-rollout}
\vspace{\subsecvspace}

A hyperscaler's ability to iterate on NIC firmware is bounded not
only by how quickly a fix can be developed and validated, but by how
quickly it can be deployed across the fleet. At Meta's scale,
firmware upgrades for server components are carried out by a
fleet-wide maintenance orchestration system that periodically visits
hosts in batches, drains them from production, applies pending
upgrades, and returns them to service. For NIC firmware, this system
provides a best-case rollout cycle of approximately 60~days for
$>$90\% fleet coverage, with a long tail that can extend
considerably further.

For a new NIC entering production, this cadence can present three
challenges. First, fbnic requires rapid firmware iteration during
early deployment, as bugs discovered in production needed fixes rolled
out in weeks, not months. Second, the initial deployment was sparse:
fbnic was present in a fraction of racks across a small number of
regions, meaning that many maintenance batches contained few or no
fbnic hosts, diluting rollout velocity. Third, the multi-host
architecture introduces a scope constraint: the standard maintenance
system upgrades firmware at the sled granularity, requiring all hosts
in a sled to be healthy and reachable. On a multi-host NIC platform
where each sled contains multiple hosts, a single host in repair
would block the NIC firmware upgrade for the entire sled, creating a
persistent tail of hosts that maintenance trains could not reach.

\paragraph{Targeted Upgrade Orchestrator.}
To address these limitations, we built a dedicated upgrade
orchestration service that complements the standard maintenance
system. The orchestrator operates
by scheduling targeted firmware upgrades as individual unplanned
maintenance events, rather than waiting for the next scheduled
maintenance batch to visit a host.

The service is organized around the concept of \emph{upgrade
campaigns}. An operator creates a campaign specifying the target
firmware version, the NIC component to upgrade, and a set of search
criteria identifying eligible hosts (e.g., hosts running a
particular firmware version in a given region). The
orchestrator then identifies matching hosts from a fleet inventory
service, batches them into phases, and executes upgrades
asynchronously. Before initiating each upgrade, it checks for
conflicting maintenance windows and verifies regional capacity availability
to avoid exceeding drain budgets. The underlying firmware flashing
and activation logic is identical to the standard maintenance
path, only the scheduling and targeting layer differs, ensuring
that the orchestrator does not introduce a separate, less-tested
upgrade code path.

One design choice was supporting upgrades scoped to the hosts
sharing a single NIC rather than the entire sled. The standard
maintenance system required all hosts on a sled to be drained for
any NIC firmware upgrade. The orchestrator reduces this to draining
only the hosts that share the target NIC, leaving the remaining
hosts on the sled in production. This directly addressed the long-tail problem:
hosts that the standard maintenance system could not reach---because
a sibling host on the same sled was in repair---could now be
upgraded individually.

To bound production impact, the orchestrator enforces a concurrency limit on the number of hosts
simultaneously undergoing upgrades in any given region, coordinated
with the capacity planning system's view of available failure
buffers.

\paragraph{Impact.}
In its first year of operation, the orchestrator complemented the
standard maintenance system across several rollout campaigns,
performing tens of thousands of firmware upgrades:

\begin{itemize}
\item \textbf{Long-tail closure:} closed a 5--10\% fleet tail
  unreachable by maintenance trains due to sibling hosts in
  repair.

\item \textbf{Accelerated rollout:} performed 35\% of all upgrades
  for one release, completing the rollout in under four weeks vs.\
  two months with trains alone.

\item \textbf{Production canaries:} enabled targeted canary
  deployments on a small percentage of production hosts before
  committing to fleet-wide rollout.
\end{itemize}

\noindent
Beyond fbnic, the orchestrator was also used to accelerate firmware
rollouts for other platform components on the same chassis, such as
retimer and baseboard controller firmware, demonstrating that the
orchestration model generalizes beyond NIC firmware.

\section{Production Incidents \& Lessons}
\label{sec:incidents}
\vspace{\subsecvspace}

Operating a first-generation custom NIC at fleet scale produces
incidents that stress not only the hardware but the entire
operational infrastructure, including tooling, diagnostics, deployment
pipelines, and cross-team coordination. We describe two production
incidents that exemplify distinct failure modes; both are
representative of a broader pattern observed across our
multi-host NIC SEV history (${\sim}14$ vendor firmware-class SEVs
over six years, median resolution 20~weeks): rare in count but
disproportionately high in operational complexity, with
infrastructure gaps acting as multipliers.

\subsection{PXE Boot Failure}
\label{subsec:pxe-sev}
\vspace{\subsecvspace}

During fbnic's initial production ramp, hosts began failing
PXE boot in increasing numbers, with approximately 3\% of the
deployed fbnic units affected and hundreds of repair tickets
opened. Because the
multi-host architecture meant that working hosts on the same sled
were held in a non-production state while their failed neighbors
were investigated, the effective capacity impact was substantially
larger than the failure count alone.

\paragraph{Root cause.}
The fbnic UEFI network driver, used only during PXE boot and not
during normal operation, depended on the UEFI network stack to
provide receive buffers, which were sized for the standard
1500-byte MTU. When a larger frame arrived, the buffer was too
small to hold it; the frame became stuck in the receive ring,
head-of-line blocking all subsequent receives. This was a day-0
firmware bug, triggered when a fleet-wide configuration change
enabled jumbo MTU (9000 bytes) for an infrastructure health
checking service, causing health check probes with payloads
exceeding 1500 bytes to reach hosts during the PXE boot sequence.

\paragraph{Debugging challenges.}
The root cause took three weeks to identify, not because the bug was
complex but because the debugging infrastructure was inadequate for
diagnosing failures during the UEFI boot phase:

\begin{itemize}
\item \textbf{No packet capture during UEFI boot.} The UEFI
  environment has no equivalent of \texttt{tcpdump}. The team could
  not see what packets the host was receiving at the time of failure.

\item \textbf{Switch-side mirroring limitations.} Setting up port
  mirroring on the rack switch to capture traffic required CLI
  commands that failed due to configuration management conflicts.
  When mirroring was finally established, the switching platform
  only supported ingress mirroring on the relevant port type---the
  response packets from the TFTP server to the host were invisible
  in the capture. The switch CPU's packet capture buffer was limited
  to 10,000 packets, insufficient to observe the full PXE boot
  transaction. Server-side capture was also unavailable because the
  TFTP server was deployed behind a load balancer.

\item \textbf{No traffic filtering on the switch.} There was no
  mechanism to filter or ACL traffic on the rack switch's downlink
  port during PXE boot, making it impossible to isolate the host's
  PXE traffic from the noise of health check probes and stale
  connections from previously drained services.
\end{itemize}

\noindent
The breakthrough came when a full console dump from a failing host
surfaced an error from the receive call indicating the buffer was
too small to handle the incoming frame; packet captures then
confirmed that all failed cases contained frames larger than 1500
bytes not present in successful boots. Deliberately injecting
jumbo-sized pings during PXE boot on a previously healthy host
reproduced the failure, confirming the root cause.

\paragraph{Recovery challenges.}
The fix, a UEFI driver update to handle jumbo frames, was
straightforward, but deploying it exposed a chicken-and-egg problem:
the standard firmware upgrade process requires PXE boot, which was
precisely the operation that was failing. Additional obstacles
compounded the recovery:

\begin{itemize}
\item \textbf{No hotfix build process.} Building an emergency firmware release with only the UEFI component changed was blocked by undefined procedures for graft-based builds, signing permissions, and RPM packaging limitations.

\item \textbf{Invisible rollout progress.} The fleet telemetry
  pipeline only reported the \emph{active} firmware version; a
  hotfix flashed to the standby partition but not yet activated was
  invisible to monitoring, preventing the team from tracking
  rollout progress.

\item \textbf{Sled-level upgrade scope.} The standard maintenance
  system upgrades firmware at sled granularity, skipping any sled
  with a host in repair. Since the PXE failure itself created
  repair tickets, affected sleds were systematically excluded from
  the upgrade that would fix them.
\end{itemize}

\noindent
The team implemented the fix through a combination of manual hotfix
installation, an interim mitigation (clamping the health checker's
TCP MSS to avoid jumbo frames on the wire), and ultimately the
targeted upgrade orchestrator (\S\ref{subsec:fw-rollout}) operating
at sub-sled granularity.

\paragraph{Lessons and follow-ups.}
A single UEFI bug surfaced over a dozen infrastructure gaps spanning
packet capture tooling, switch mirroring, firmware build and deploy
pipelines, telemetry visibility, and test coverage. Each gap was
independently reasonable---no one had anticipated needing
\texttt{tcpdump} during UEFI boot---but together they turned a one-week fix into a six-week recovery, a factor of 2--3$\times$ faster than the 10--20~weeks that comparable vendor PXE/UEFI bugs required. The incident directly
motivated adding PXE boot with jumbo frame injection to the CI
pipeline's test suite (\S\ref{sec:testing}), building a defined
hotfix firmware process, adding standby firmware version reporting
to the telemetry pipeline, and developing the sub-sled upgrade
capability in the targeted upgrade orchestrator.

\subsection{Silent Memory Corruption}
\label{subsec:bitflip-sev}
\vspace{\subsecvspace}

A storage service experienced silent data corruption
exclusively on hosts with an fbnic NIC, detected through automated
alerts indicating service panics. Early hypotheses explored firmware
slowdowns and race conditions in the host mailbox interface, as the
team found correlations between missed firmware heartbeats and
corruption events.

\paragraph{Root cause.}
The root cause was a race condition in the NIC firmware's DMA
address translation logic. The firmware used a fixed base
address (\texttt{0x200000000}) representing its host memory window.
For each DMA write to host memory, the firmware programmed an
outbound iATU (internal Address Translation Unit) entry to translate
writes from this internal address to the actual host destination,
performed the write, then disabled the iATU. However, PCIe writes
are ``posted,'' meaning the firmware had no visibility into whether
data had actually left the on-chip AXI-to-PCIe interface FIFO. Under
backpressure from the PCIe root complex---exacerbated by the CPU
power management behavior described in
\S\ref{subsec:pcie-latency}---writes could remain buffered longer
than expected. If the firmware disabled the iATU before the write
exited the FIFO, the transaction proceeded without translation,
landing at the raw internal address instead of the intended
destination. Because IOMMU was in passthrough mode for many
workloads, this raw address could land in valid host memory rather
than being caught as a fault, causing silent corruption.


\paragraph{Timeline.} The bug surfaced in production as a rare event. The first two to three weeks went to triage: convincing ourselves the NIC was at fault and building a reliable reproduction, achieved by inducing PCIe backpressure with the IOMMU enabled so the silent writes surfaced as catchable faults.
The reproduction unblocked two parallel efforts. Even once qualified, the firmware fix would take roughly 4-6 more weeks after qualification to reach the fleet, too long to leave production exposed, so we deployed a stopgap in parallel: marking the affected memory page offline, which rendered stray writes harmless and removed production impact. The permanent firmware fix round-robined all 16 outbound iATU entries instead of reusing one, disabling each only after a read confirmed its writes had flushed, and cleared two-stage CI qualification about two weeks later, five weeks after
the SEV opened. With impact already contained, the qualified firmware
rolled out through two scheduled deployment freezes, completing fleet-wide roughly six weeks afterward.

\paragraph{Lessons.}
This incident illustrates the value of owning the firmware stack.
The root cause---a subtle interaction between PCIe posted-write
semantics and the iATU disable timing---would be extremely
difficult for a vendor to diagnose remotely. Because the firmware
was developed in-house, the team could reproduce the bug, develop
the mitigation and permanent fix, and deploy it through the same
CI pipeline used for all firmware changes. For comparison, a
firmware bug of similar severity on a vendor multi-host NIC required
52~weeks to resolve (\S\ref{subsec:pause-cascade}).

In hindsight, both incidents point to the same lesson: the UEFI
boot environment and firmware DMA paths require the same level of
testing rigor as the data-path driver. We now test PXE boot under
jumbo frame injection and validate iATU behavior under PCIe
backpressure as part of every firmware release.

\vspace{-10pt}
\section{Evaluation}
\label{sec:evaluation}
\vspace{\subsecvspace}
We evaluate the operational
infrastructure described in this paper across three dimensions:
fleet reliability, firmware iteration velocity, and performance
parity with vendor NICs.

\subsection{Fleet Reliability}
\label{subsec:eval-reliability}
\vspace{\subsecvspace}

\begin{figure}[t]
\centering
\scriptsize
\begin{tikzpicture}
\begin{axis}[
    width=0.95\columnwidth,
    height=0.45\columnwidth,
    xlabel={},
    ylabel={Unplanned UU (\%)},
    xmin=0.5, xmax=10.5,
    ymin=0, ymax=10,
    xtick={1,2,3,4,5,6,7,8,9,10},
    xticklabels={M1,M2,M3,M4,M5,M6,M7,M8,M9,M10},
    x tick label style={rotate=45, anchor=east},
    ymajorgrids=true,
    grid style={gray!20},
    legend style={draw=none, fill=white, fill opacity=0.8,
      at={(0.98,0.98)}, anchor=north east, font=\scriptsize},
    legend cell align={left},
]
\addplot[thick, mark=*, blue, mark size=1.5pt] coordinates {
    (1,9.11) (2,5.88) (3,6.76) (4,5.94) (5,2.54) (6,0.99) (7,0.74) (8,1.25) (9,1.95) (10,1.44)
};
\addlegendentry{fbnic T1}
\addplot[thick, mark=square*, red!70!black, mark size=1.5pt] coordinates {
    (1,7.52) (2,2.54) (3,4.27) (4,1.17) (5,1.24) (6,0.85) (7,0.75) (8,0.79) (9,0.80) (10,1.14)
};
\addlegendentry{Vendor T1}
\node[font=\tiny, anchor=south] at (axis cs:7,0.74) {0.74\%};
\node[font=\tiny, anchor=north] at (axis cs:7,0.75) {0.75\%};
\end{axis}
\end{tikzpicture}

\vspace{-2mm}

\begin{tikzpicture}
\begin{axis}[
    width=0.95\columnwidth,
    height=0.35\columnwidth,
    xlabel={Month},
    ylabel={Fleet size ($\times$ M1 baseline)},
    xmin=0.5, xmax=10.5,
    ymin=0, ymax=140,
    xtick={1,2,3,4,5,6,7,8,9,10},
    xticklabels={M1,M2,M3,M4,M5,M6,M7,M8,M9,M10},
    x tick label style={rotate=45, anchor=east},
    ytick={1,50,100,130},
    ymajorgrids=true,
    grid style={gray!20},
    area style,
]
\addplot[thick, mark=*, fill=blue!15, draw=blue!70, mark size=1.5pt] coordinates {
    (1,1) (2,2.7) (3,8.7) (4,13.7) (5,17.8) (6,20.4) (7,27.3) (8,62.1) (9,82.3) (10,130)
};
\end{axis}
\end{tikzpicture}
\caption{Top: Unplanned unavailability for fbnic vs.\ vendor NIC
on the same platform (T1 tier). fbnic UU improved 12$\times$
from 9.11\% to 0.74\%, reaching parity with the vendor baseline
by month M7. Bottom: fbnic fleet grew 130$\times$ over the same
10-month window, with M8--M9 contributing the largest acceleration.}
\label{fig:uu-trend}
\end{figure}
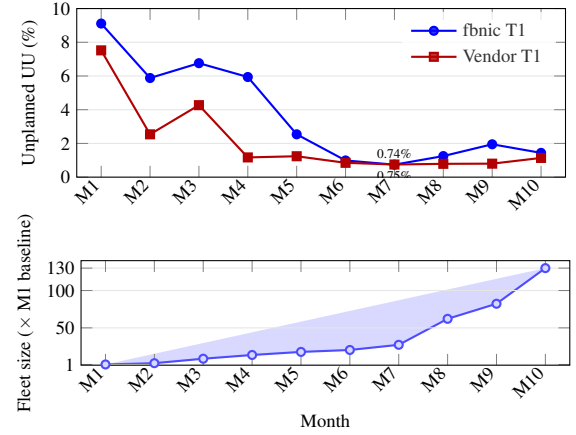

\paragraph{Unavailability improvement.}
T1~fbnic unplanned unavailability (UU) improved from 9.11\% to
0.74\%, a 12$\times$ reduction over six months (M1 to M7),
reaching parity with the vendor NIC baseline on the same platform
(Figure~\ref{fig:uu-trend}). These are all-cause UU figures, inclusive of the two issues discussed below.The high initial UU was driven by the PXE boot incident (\S\ref{subsec:pxe-sev}); the improvement reflects the PXE fix reaching fleet coverage, repair action plan augmentation, and firmware rollouts. A subsequent regression in
months M8--M9 correlated with the fleet tripling in size and
a rack-switch interoperability issue resolved by a firmware update.

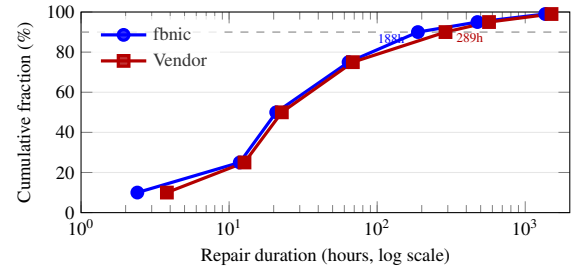
\begin{figure}[t]
\centering
\scriptsize
\begin{tikzpicture}
\begin{axis}[
    width=0.95\columnwidth,
    height=0.5\columnwidth,
    xlabel={Repair duration (hours, log scale)},
    ylabel={Cumulative fraction (\%)},
    xmode=log,
    xmin=1, xmax=2000,
    ymin=0, ymax=100,
    ytick={0,20,40,60,80,100},
    ymajorgrids=true,
    grid style={gray!20},
    legend style={draw=none, fill=white, fill opacity=0.8,
      at={(0.02,0.98)}, anchor=north west, font=\scriptsize},
    legend cell align={left},
]
\addplot[very thick, mark=*, blue, mark size=2pt] coordinates {
    (2.4,10) (11.8,25) (20.8,50) (64.1,75) (188.4,90) (475.6,95) (1364.3,99)
};
\addlegendentry{fbnic}
\addplot[very thick, mark=square*, red!70!black, mark size=2pt] coordinates {
    (3.8,10) (12.7,25) (22.7,50) (68.5,75) (289.4,90) (569.6,95) (1505.1,99)
};
\addlegendentry{Vendor}
\draw[dashed, gray] (axis cs:1,90) -- (axis cs:2000,90);
\node[font=\tiny, blue, anchor=east] at (axis cs:170,87) {188h};
\node[font=\tiny, red!70!black, anchor=west] at (axis cs:310,87) {289h};
\end{axis}
\end{tikzpicture}
\caption{CDF of NIC repair duration (TTR) on the same platform.
fbnic repairs are faster at every percentile. At p90, fbnic
completes in 188~hours vs.\ 289~hours for the vendor NIC.}
\label{fig:mttr-cdf}
\end{figure}

\paragraph{Repair effectiveness.}
Table~\ref{tab:repair-compare} compares repair outcomes for
fbnic and vendor NICs on the same platform over ten months.
fbnic demonstrates advantages across every serviceability
dimension: 37\% lower mean time to repair, 2.3$\times$
fewer NIC swaps, 63\% higher reseat resolution rate (reseats that
actually resolve the issue vs.\ requiring escalation), and 57\%
higher diagnostic accuracy (automated diagnosis correctly
identifying the root cause). Figure~\ref{fig:mttr-cdf} shows the
time-to-repair (TTR) CDF: fbnic repairs are faster at every percentile, with
the gap widening at the tail (p90: 188~hours vs.\ 289~hours).

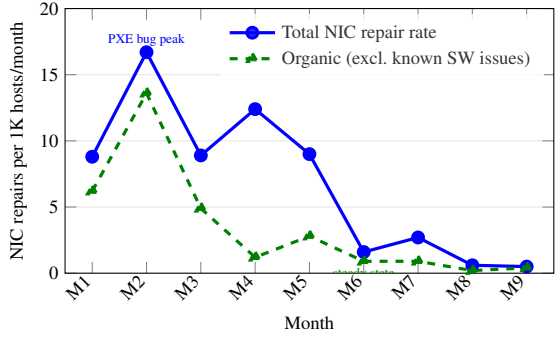
\begin{figure}[t]
\centering
\scriptsize
\begin{tikzpicture}
\begin{axis}[
    width=0.95\columnwidth,
    height=0.6\columnwidth,
    xlabel={Month},
    ylabel={NIC repairs per 1K hosts/month},
    xmin=0.5, xmax=9.5,
    ymin=0, ymax=20,
    xtick={1,2,3,4,5,6,7,8,9},
    xticklabels={M1,M2,M3,M4,M5,M6,M7,M8,M9},
    x tick label style={rotate=45, anchor=east},
    ymajorgrids=true,
    grid style={gray!20},
    legend style={draw=none, fill=white, fill opacity=0.8,
      at={(0.98,0.98)}, anchor=north east, font=\scriptsize},
    legend cell align={left},
]
\addplot[very thick, mark=*, blue, mark size=2pt] coordinates {
    (1,8.8) (2,16.7) (3,8.9) (4,12.4) (5,9.0) (6,1.6) (7,2.7) (8,0.6) (9,0.5)
};
\addlegendentry{Total NIC repair rate}
\addplot[very thick, mark=triangle*, green!50!black, mark size=2pt, dashed] coordinates {
    (1,6.2) (2,13.6) (3,4.9) (4,1.2) (5,2.8) (6,0.9) (7,0.9) (8,0.2) (9,0.4)
};
\addlegendentry{Organic (excl.\ known SW issues)}
\node[font=\tiny, blue, anchor=south] at (axis cs:2,16.7) {PXE bug peak};
\node[font=\tiny, green!50!black, anchor=north] at (axis cs:6,0.9) {steady state};
\end{axis}
\end{tikzpicture}
\caption{Monthly NIC repair rate per 1,000 hosts. Two known
software/interop issues (PXE boot bug and rack-switch PHY
interaction) accounted for ${\sim}$60\% of repairs. The organic
rate (excluding these) stabilized below 1.0 per 1K hosts/month,
comparable to vendor NICs on the same platform.}
\label{fig:failure-rate}
\end{figure}

Two known software and interoperability issues, the PXE boot bug
(\S\ref{subsec:pxe-sev}) and a rack-switch PHY interaction,
accounted for approximately 60\% of all fbnic NIC repair events
during the evaluation period. Neither reflects inherent hardware
reliability; both were identified and resolved within 6--10~weeks
through the in-house firmware pipeline. Excluding these, the
organic NIC failure rate stabilized at 0.2--0.9~per 1,000~hosts
per month by the final months of the observation window,
comparable to the vendor NIC baseline
(Figure~\ref{fig:failure-rate}). The TTR and swap-rate figures in Table~\ref{tab:repair-compare} are reported over all repair events, on the same inclusive basis as the vendor NIC.

\begin{table}[t]
\centering
\small
\begin{tabular}{lrr}
\hline
\textbf{Metric} & \textbf{fbnic} & \textbf{Vendor} \\
\hline
Avg TTR (hours)         & \textbf{125.5} & 197.9 \\
P90 TTR (hours)         & \textbf{188}   & 289 \\
Swap rate (\%)           & \textbf{2.9}   & 6.7 \\
Reseat resolve rate (\%) & \textbf{63.2}  & 38.8 \\
Diagnostic accuracy (\%) & \textbf{49.2}  & 31.4 \\
\hline
\end{tabular}
\caption{Repair effectiveness on the same platform over a
10-month observation window. fbnic is repaired faster, requires
fewer hardware swaps, and achieves higher diagnostic precision.}
\label{tab:repair-compare}
\end{table}

\paragraph{Performance parity.}
Table~\ref{tab:perf-parity} compares NIC health metrics between
fbnic and vendor NICs running \emph{the same production
services} on the same platform, controlling for workload
differences. fbnic and the vendor NIC use different flow control
strategies: fbnic sends TX pause frames to the switch to prevent
buffer overflow (\S\ref{subsec:pcie-latency}), while the vendor
NIC drops packets when its larger buffer is exhausted. We show
both RSW in-pause (pause frames received by the switch from the
NIC) and RX out-of-buffer to present this tradeoff transparently.
Despite sending pause frames, fbnic achieves comparable or
lower TCP retransmit rates across all measured workloads, and
switch-side FEC uncorrectable errors are 3$\times$ lower.

\begin{table}[t]
\centering
\small
\begin{tabular}{llrrrr}
\hline
\textbf{Service} & \textbf{NIC} & \textbf{TCP} & \textbf{RSW} & \textbf{RX} & \textbf{RSW FEC} \\
 & & \textbf{rxmit} & \textbf{pause} & \textbf{OOB} & \textbf{uncorr.} \\
\hline
\multirow{2}{*}{Web}   & fbnic & \textbf{0.44} & 0.006 & \textbf{0.0} & \textbf{0.33} \\
                       & Vendor & 0.46 & 0.0 & 8.3 & 1.05 \\
\hline
\multirow{2}{*}{Cache} & fbnic & \textbf{0.19} & 105.9 & \textbf{0.0} & \textbf{0.28} \\
                       & Vendor & 0.24 & 0.0 & 8,958 & 0.95 \\
\hline
\multirow{2}{*}{Ads}   & fbnic & \textbf{0.22} & 2.69 & \textbf{0.0} & \textbf{0.41} \\
                       & Vendor & 0.22 & 0.0 & 0.05 & 1.12 \\
\hline
\end{tabular}
\caption{NIC health metrics for the same services on the same
platform (T1, end of observation window). Lower is better. fbnic uses TX pause
to prevent buffer drops (RSW pause column); the vendor NIC drops
packets instead (RX OOB column). Despite pausing, fbnic matches
or beats vendor TCP retransmit rates.}
\label{tab:perf-parity}
\end{table}

\vspace{-10pt}
\subsection{Firmware Iteration Velocity}
\label{subsec:eval-velocity}
\vspace{\subsecvspace}

\paragraph{Bug-fix-to-fleet cycle.}
The silent memory corruption incident (\S\ref{subsec:bitflip-sev})
demonstrates the end-to-end firmware fix cycle: from SEV
identification to CI-qualified firmware in approximately
five weeks (reproduction in 2~weeks, firmware fix in 1~week, CI
qualification in 2~weeks). Fleet-wide deployment completed
approximately six weeks after scheduled deployment freezes ended.
By comparison, vendor multi-host NIC firmware bugs from our SEV
history typically required around 5~months to resolve, with
extreme cases approaching a year (e.g., 52~weeks for the
pause-frame issue in \S\ref{subsec:pause-cascade}). Even
the fastest closures were bounded by the 7--10~week validation
pipeline (vendor, ODM, and internal qualifications).

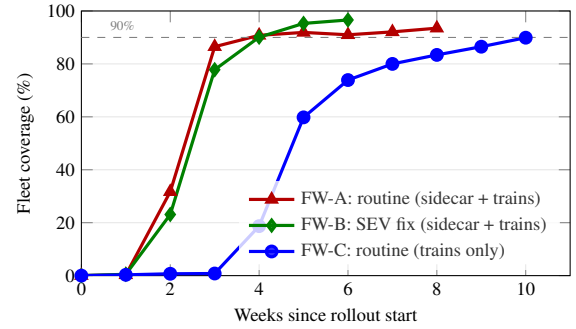
\begin{figure}[t]
\centering
\scriptsize
\begin{tikzpicture}
\begin{axis}[
    width=0.95\columnwidth,
    height=0.6\columnwidth,
    xlabel={Weeks since rollout start},
    ylabel={Fleet coverage (\%)},
    xmin=0, xmax=11,
    ymin=0, ymax=100,
    ytick={0,20,40,60,80,100},
    ymajorgrids=true,
    grid style={gray!20},
    legend style={draw=none, fill=white, fill opacity=0.8,
      at={(0.98,0.02)}, anchor=south east, font=\scriptsize},
    legend cell align={left},
]
\addplot[very thick, mark=triangle*, red!70!black, mark size=2pt] coordinates {
    (0,0) (1,0.3) (2,31.7) (3,86.5) (4,90.8) (5,91.9) (6,91.0) (7,92.1) (8,93.5)
};
\addlegendentry{FW-A: routine (sidecar + trains)}
\addplot[very thick, mark=diamond*, green!50!black, mark size=2pt] coordinates {
    (0,0) (1,0.5) (2,23.1) (3,77.8) (4,90.1) (5,95.3) (6,96.6)
};
\addlegendentry{FW-B: SEV fix (sidecar + trains)}
\addplot[very thick, mark=*, blue, mark size=2pt] coordinates {
    (0,0) (1,0.3) (2,0.7) (3,0.8) (4,18.7) (5,59.8) (6,73.9) (7,80.0) (8,83.4) (9,86.5) (10,89.9)
};
\addlegendentry{FW-C: routine (trains only)}
\draw[dashed, gray] (axis cs:0,90) -- (axis cs:11,90);
\node[font=\tiny, gray, anchor=south west] at (axis cs:0.5,90) {90\%};
\end{axis}
\end{tikzpicture}
\caption{Firmware rollout velocity comparison. With the targeted
upgrade orchestrator performing 35--40\% of upgrades (FW-A and
FW-B), 90\% coverage was reached in 4~weeks. The critical bitflip
fix (FW-B) reached 96\% in 5~weeks. Without the orchestrator
(FW-C, maintenance trains only), the same coverage took
10~weeks.}
\label{fig:fw-rollout}
\end{figure}

\paragraph{Firmware rollout.}
The targeted upgrade orchestrator (\S\ref{subsec:fw-rollout})
reduced fleet-wide firmware rollout from over 60~days for 90\%
coverage (and over 120~days for the long tail) using maintenance
trains alone, to under 4~weeks
(Figure~\ref{fig:fw-rollout}). The orchestrator's main
contribution was reaching the 5--10\% fleet tail per release
that maintenance trains could not, because sibling hosts were in
repair. For later releases, including the critical fix for the
memory corruption incident (\S\ref{subsec:bitflip-sev}), the
orchestrator handled 35--40\% of upgrades.

\vspace{-10pt}
\section{Related Work}
\label{sec:related}
\vspace{\subsecvspace}

\paragraph{NIC resource isolation.}
Prior work on multi-tenant NIC isolation has focused on software
mechanisms within shared hardware. Loom~\cite{Stephens2019LoomFA}
proposes flexible NIC packet scheduling to prevent head-of-line
blocking. Kong et al.~\cite{kong-nsdi23} characterize RDMA
microarchitecture resource contention and demonstrate that
host-level isolation requires understanding NIC-internal resource
sharing. FairNIC~\cite{GrantFairNic} implements programmable
isolation policies on SmartNICs. Recent work argues that the NIC
should be more tightly integrated with the OS to reduce firmware
complexity~\cite{nic-part-of-os}, an observation consistent with
fbnic's driver-first architecture (\S\ref{subsec:driver-first}).
These approaches work within the constraints of shared-NIC
architectures; fbnic takes the complementary approach of
eliminating shared resources through physical isolation, trading
NIC complexity for operational simplicity.

\paragraph{Datacenter network observability.}
Network fault localization systems such as
NetBouncer~\cite{netbouncer}, NetNorad~\cite{netnorad},
Flock~\cite{flock}, and 007~\cite{007} diagnose faults through
end-to-end probing or host-observed retransmissions. Fleet-scale
host networking systems~\cite{pingmesh, snap, accelnet} monitor NIC
health but treat the attached switch as a separate diagnostic
domain. Meta's Evolvable Network Telemetry~\cite{evolvable-telemetry}
provides flexible switch-side telemetry but does not collect
host-side NIC hardware counters. In-band network
telemetry~\cite{hpcc, grayint} embeds switch metadata into packets
for congestion control. Our observability pipeline
(\S\ref{subsec:observability}) extends the existing host telemetry
infrastructure with NIC-specific hardware counters co-located with
matched switch port counters
in a single per-host sample, enabling cross-layer fault attribution
without requiring in-band telemetry support from the switching
platform.

\paragraph{Custom hardware operations at scale.}
Azure AccelNet~\cite{accelnet} describes deploying FPGA-based
SmartNICs across millions of servers but focuses on the networking
architecture rather than the operational lifecycle. Google's
TPU~\cite{jouppi-tpu} and Meta's MTIA~\cite{freya} describe custom
accelerator silicon but emphasize workload performance over fleet
deployment, testing, and repair infrastructure. Google's fleet
management for multi-node servers~\cite{google-fleet} addresses
firmware and security management at scale but does not describe
NIC-specific upgrade orchestration or isolation-aware repair
scoping. Prior work on datacenter repair
automation~\cite{repair-prediction} predicts remediation actions
for hardware failures but does not address NIC-specific scoping
based on physical isolation. To our knowledge, this paper is the
first to describe the full operational stack---CI pipeline,
observability, firmware rollout, scoped repair, and production
incident response---for a custom NIC deployed at hyperscale.

\vspace{-10pt}
\section{Conclusion}
\label{sec:conclusion}
\vspace{\subsecvspace}

\begin{table}[t]
\centering
\small
\begin{tabular}{lccc}
\hline
\textbf{Contribution} & \textbf{Phys.} & \textbf{In-house} & \textbf{Portable} \\
                      & \textbf{isol.} & \textbf{FW/driver} & \textbf{(commodity)} \\
\hline
Isolation architecture         & \checkmark &            &            \\
Scoped repair automation   & \checkmark &            &            \\
Driver-first control plane   &            & \checkmark &            \\
Owning the fix cycle       &            & \checkmark &            \\
Observability pipeline     &            &            & \checkmark \\
Hardware-in-the-loop CI            &            &            & \checkmark \\
Diagnostic log collection  &            &            & \checkmark \\
Upgrade orchestrator         &            &            & \checkmark \\
\hline
\end{tabular}
\caption{Which contributions require custom hardware, in-house
firmware/driver ownership, or neither. Four of eight are portable
fleet-operational practices usable on commodity hardware.}
\label{tab:contribution-scope}
\end{table}

This paper described the operational infrastructure required to
deploy and operate fbnic, a custom multi-host NIC, at fleet
scale. The key insight is that physical isolation at the hardware
level simplifies the entire operational stack: fbnic's per-host
isolation eliminated a cascade of software workarounds that had
accumulated over seven years, and enabled scoped repair, sub-sled
firmware upgrades, and per-host fault attribution that were
impossible with shared-firmware vendor architectures.

Not every technique in this paper depends on custom silicon.
Table~\ref{tab:contribution-scope} classifies our contributions by what
each requires. Only the isolation architecture and scoped repair depend on
the custom hardware; two more depend on owning the firmware and driver
stack; the remaining four are general fleet-operational practices
applicable to any operator running commodity hardware. Our observability
pipeline, for instance, is already deployed across both in-house and
vendor NICs fleet-wide.

Quantitatively, fbnic achieved a 12$\times$ reduction in
unplanned unavailability, a 2.3$\times$ lower hardware swap rate on
the same platform, and a firmware fix cycle of five weeks compared
to around 5~months (with extreme cases approaching a year) for
vendor NICs. Looking ahead, the operational
patterns described here are being adapted for other custom silicon
programs including Meta's MTIA accelerator~\cite{freya}. The number of hosts sharing a NIC is itself a tunable parameter that decreases as per-host bandwidth grows, so the isolated-slice architecture adapts to rising bandwidth demands rather than becoming obsolete. The lessons from fbnic's
first year in production apply to any organization bringing custom
hardware from lab to fleet for the first time.

\newpage

{\sloppy
\bibliographystyle{plain}
\bibliography{references}
}

\end{document}